\documentclass[letterpaper,10 pt,conference]{ieeeconf}
\IEEEoverridecommandlockouts 
\usepackage{amsmath}
\usepackage{algorithmic}
\usepackage{array}
\usepackage{stfloats}
\usepackage{url}
\usepackage{cite}
\usepackage{graphicx}
\usepackage{makecell}
\usepackage{mathtools}
\usepackage{threeparttable}
\usepackage{amssymb}
\usepackage{bm}
\usepackage{array} 
\usepackage{float}
\usepackage{subfigure}
\usepackage{gensymb}
\usepackage{textcomp}
\usepackage{wasysym}
\usepackage[table,dvipsnames]{xcolor}
\usepackage{hyperref}
\usepackage{booktabs}
\usepackage{ragged2e}
\usepackage{blkarray,bigstrut}
\usepackage{multirow}
\usepackage{colortbl,hhline}
\usepackage{hyphsubst}
\usepackage{makecell}
\HyphSubstLet{english}{usenglishmax}
\usepackage[english]{babel}
\usepackage{color,soul}
\usepackage{etoolbox}
\makeatletter
\patchcmd{\@makecaption}
  {\scshape}
  {}
  {}
  {}
\makeatother

\title{\LARGE \bf
Adapting SAM for Surgical Instrument Tracking and Segmentation in Endoscopic Submucosal Dissection Videos
}

\author{Jieming Yu, Long Bai$^\dagger$, Guankun Wang, An Wang, Xiaoxiao Yang, Huxin Gao, \\ Hongliang Ren$^*$, \emph{Senior Member, IEEE}
\thanks{*This work was supported by Hong Kong Research Grants Council (RGC) Collaborative Research Fund (CRF) C4026-21GF, General Research Fund (GRF) 14203323, GRF 14216022, GRF 14211420, NSFC/RGC Joint Research Scheme N\_CUHK420/22, Shenzhen-Hong Kong-Macau Technology Research Programme (Type C) STIC Grant 202108233000303. (Corresponding author: H. Ren, hlren@ieee.org)}%
\thanks{$^\dagger$Project Lead.}
\thanks{J. Yu, L. Bai, G. Wang, A. Wang, H. Gao, and H. Ren are with The Chinese University of Hong Kong, Hong Kong, China. X. Yang is with Qilu Hospital of Shandong University, Jinan, China. H. Ren is also with the Shenzhen Research Institute, The Chinese University of Hong Kong, and the Department of Biomedical Engineering, National University of Singapore.}
}

\begin{document}

\maketitle
\thispagestyle{empty}
\pagestyle{empty}

\begin{abstract}
The precise tracking and segmentation of surgical instruments have led to a remarkable enhancement in the efficiency of surgical procedures. However, the challenge lies in achieving accurate segmentation of surgical instruments while minimizing the need for manual annotation and reducing the time required for the segmentation process. To tackle this, we propose a novel framework for surgical instrument segmentation and tracking. Specifically, with a tiny subset of frames for segmentation, we ensure accurate segmentation across the entire surgical video. Our method adopts a two-stage approach to efficiently segment videos. Initially, we utilize the Segment-Anything (SAM) model, which has been fine-tuned using the Low-Rank Adaptation (LoRA) on the EndoVis17 Dataset. The fine-tuned SAM model is applied to segment the initial frames of the video accurately. Subsequently, we deploy the XMem++ tracking algorithm to follow the annotated frames, thereby facilitating the segmentation of the entire video sequence. This workflow enables us to precisely segment and track objects within the video. Through extensive evaluation of the in-distribution dataset (EndoVis17) and the out-of-distribution datasets (EndoVis18 \& the endoscopic submucosal dissection surgery (ESD) dataset), our framework demonstrates exceptional accuracy and robustness, thus showcasing its potential to advance the automated robotic-assisted surgery.
\end{abstract}

\section{Introduction}\label{sec1}

In robot-assisted surgery, accurate surgical instrument tracking and segmentation can provide valuable information about the position, movement, and interactions of surgical tools during the surgical procedure\cite{wang2023sam,bai2023surgical,8648150}. 
Recently, Segment Anything Model (SAM) \cite{kirillov2023segany} has demonstrated remarkable results on its capabilities of zero-shot generalization in natural image segmentation, by utilizing input prompts such as points and bounding boxes. However, SAM encounters obstacles in its application to surgical tasks because of the limited availability of surgical data during SAM pre-training and the significant gap between natural objects and surgical targets\cite{he2023computervision}. Moreover, its direct application to videos reveals challenges in maintaining temporal correspondence, resulting in less impressive performance\cite{yang2023track}. Additionally, SAM's interactive nature requires explicit prompts for each image, introducing a time-consuming and labor-intensive process for labeling large datasets.

Given that even the smallest version (ViT-B) of SAM also encompasses 91M parameters, the feasibility of conducting full fine-tuning for SAM becomes highly impractical. Therefore, we employ Low-Rank Adaptation (LoRA) \cite{hu2021lora} to adapt the model to the surgical domain. 
Furthermore, to tackle the issue of SAM's suboptimal performance in video applications, we have integrated a state-of-the-art semi-supervised video object segmentation (SSVOS) model, named XMem++~\cite{bekuzarov2023xmem}. 
By integrating the fine-tuned SAM  with XMem++~\cite{bekuzarov2023xmem}, our framework achieves outstanding segmentation and tracking results. The detailed procedure is presented in Section~\ref{methodology}. 


Moreover, we also introduce a novel surgical video dataset named ESD (Endoscopic Submucosal Dissection)~\cite{bai2024ossar,gao2023transendoscopic,yang2023novel}. This dataset comprises 3 surgical videos with 300 frames, specifically from ESD surgery. The instrument and organ segmentation masks are annotated manually. By including this dataset, we provide a standardized benchmark for evaluating the performance of various algorithms in surgical video analysis. 
We assess the effectiveness of our framework by evaluating its performance on publicly available datasets and our ESD dataset. Our evaluation results showcase exceptional segmentation outcomes, further highlighting our impressive performance.

\begin{figure}[t!]
    \centering
    \includegraphics[width=3.3in, trim=0 5 0 0]{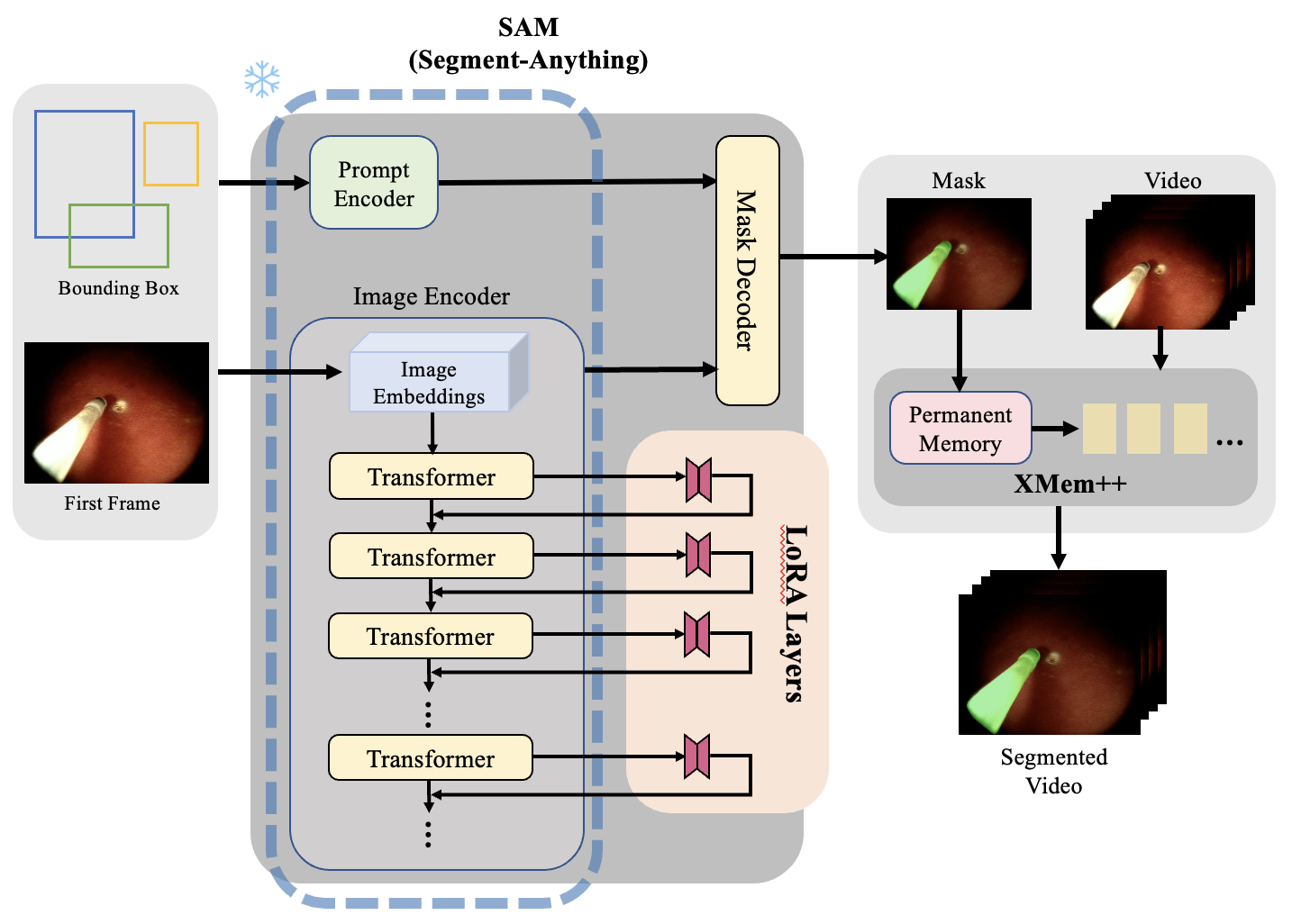}
    \label{Figure1}
    \caption{Pipeline of our proposed solution. LoRA layers are added to the transformer blocks to fine-tune the SAM model and the output mask is sent to XMem++~\cite{bekuzarov2023xmem} to track the whole video. }
\end{figure}

\section{Methodologies}
\label{methodology}
Figure~\ref{Figure1} illustrates the overall architecture of our solution. During the training phase, we employ LoRA layers to efficiently fine-tune the SAM model using the EndoVis17 dataset~\cite{cui2024surgical}. Subsequently, the pre-trained XMem++~\cite{bekuzarov2023xmem} model tracks the annotations in frames from the fine-tuned SAM model, thereby facilitating segmentation for the entire surgical video. During the inference phase, the SAM model shall segment the first several frames of the video sequence, and the XMem++~\cite{bekuzarov2023xmem} shall segment the remaining video by following the segmentation masks from SAM.

During the SAM fine-tuning process, we employ the public EndoVis17 dataset~\cite{allan2019endovis17} as our training set, comprising 1800 images. We select the ViT-B encoder due to its minimal parameter count relative to its larger counterparts. Within the SAM model, we freeze the training parameters of both the prompt and image encoders. To tailor the model for surgical scene segmentation, we introduce an additional LoRA layer to each ViT block in the image encoder. Empirically, we set a rank of 512, which offers the optimal adaptation within the LoRA matrices. The SAM model's prompt encoder encodes the bounding box prompt for the surgical instrument, which, when fused with the image encoding, generates the composite image embedding. The mask decoder then interprets this embedding to produce the image segmentation. We set 10 epochs for the fine-tuning stage.

For comprehensive segmentation across the surgical video, we employ the pre-trained XMem++~\cite{bekuzarov2023xmem} model to track the remaining video frames, using the segmentation masks from the refined SAM model. XMem++~\cite{bekuzarov2023xmem} introduces a cutting-edge approach to semi-supervised video object segmentation by innovating a permanent memory module within memory-based methodologies. This novel strategy delivers consistent segmentation outcomes while significantly reducing the need for frame annotations. By capitalizing on the segmentation output from our refined SAM model, we effectively eliminate the requirement for manual frame annotations in the original XMem++~\cite{bekuzarov2023xmem} method, further enhancing process automation.

\section{Experiments} \label{experiment}

\begin{table}[!t]
\caption{Evaluation results of original SAM and fine-tuned SAM on EndoVis17, EndoVis18, and ESD datasets.}
\centering
\resizebox{0.45\textwidth}{!}{
\begin{tabular}{c|c|ccc}
\hline 
Dataset                    & Model          & mIoU & mAcc & mDice \\ \hline
\multirow{2}{*}{EndoVis17} & Original SAM   & 40.29    & 81.08    & 50.17     \\
                           & Fine-tuned SAM & 91.38    & 98.96    & 95.06     \\ \hline
\multirow{2}{*}{EndoVis18} & Original SAM   & 32.99    & 73.52    & 42.04     \\
                           & Fine-tuned SAM & 85.28    & 97.96    & 90.21     \\ \hline
\multirow{2}{*}{ESD}       & Original SAM   & 79.67    & 97.73    & 87.66     \\
                           & Fine-tuned SAM & 82.56    & 97.78    & 89.88     \\ \hline

\end{tabular}} 
\label{Table1}
\end{table}

The performance of the fine-tuned model was evaluated on the validation sets of EndoVis17~\cite{allan2019endovis17}, EndoVis18~\cite{allan2020endovis18}, and the whole ESD dataset, with comparisons to the baseline SAM model. Our analysis, as summarized in Table~\ref{Table1}, demonstrates the fine-tuned model's marked superiority over the original SAM across all examined datasets. Concretely, with respect to EndoVis17, our fine-tuned model achieves remarkable improvements with a 51.38\% increase in mean IoU (mIoU), a 17.88\% enhancement in mean accuracy (mAcc), and a 44.89\% uplift in mean Dice coefficient (mDice). For EndoVis18, the model exhibits a significant advancement with 52.29\% in mIoU, 24.44\% in mAccuracy, and 48.17\% in mDice coefficient. Regarding the ESD dataset, the model also registers gains of 2.89\% in mIoU, 0.05\% in mAccuracy, and 2.22\% in mDice coefficient. These results underscore the fine-tuned model's robustness and its excellent performance in endoscopic surgical datasets across a spectrum of evaluation metrics.
Furthermore, we integrate the fine-tuned model with the pre-trained XMem++~\cite{bekuzarov2023xmem} model. Utilizing the fine-tuned model to generate the initial frame mask for the ESD dataset, and leveraging XMem++~\cite{bekuzarov2023xmem} for extensive video tracking, we achieve high-precision video segmentation. The comparative evaluation against the Track Anything benchmark, detailed in Table~\ref{Table2}, highlights our framework's exemplary performance.

\begin{figure}[t!]
    \centering
    \includegraphics[width=3.45in, trim=0 15 0 0]{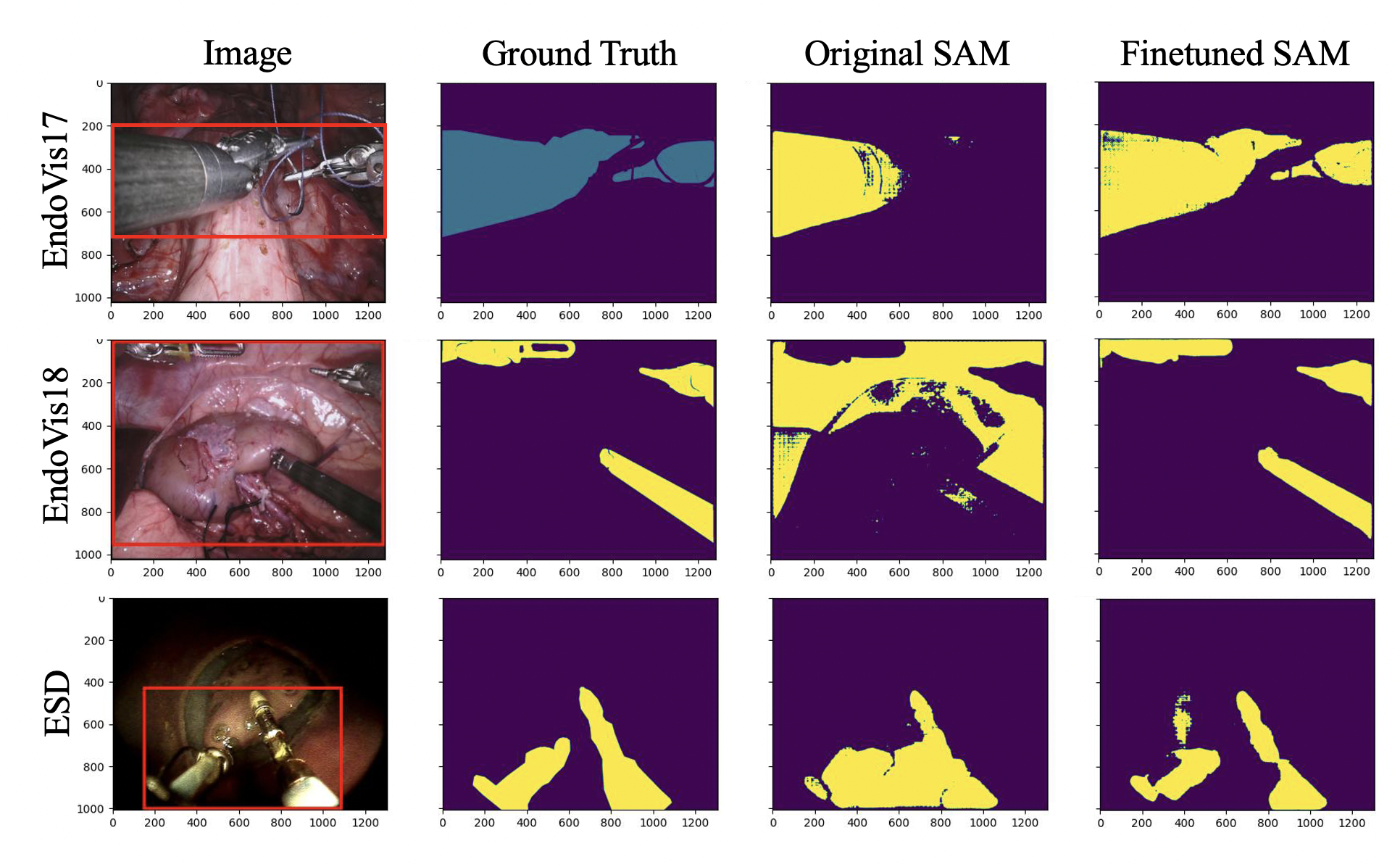}
    \caption{Visualization examples of the fine-tuned SAM on the validation set of EndoVis17, EndoVis18, and ESD datasets.}
    \label{Figure2}
\end{figure}

\begin{table}[]
\caption{Evaluation results of Fine-tuned SAM with XMem++~\cite{bekuzarov2023xmem} and Track Anything~\cite{yang2023track} on the ESD video datasets.}
\centering
\resizebox{0.47\textwidth}{!}{
\begin{tabular}{c|c|ccc}
\hline 
\multicolumn{1}{c|}{Model}          & \multicolumn{1}{c}{mIoU}        & \multicolumn{1}{c}{mAcc}   & mDice       \\ \hline
\multicolumn{1}{c|}{Track Anything~\cite{yang2023track}} & \multicolumn{1}{c}{86.75}          & \multicolumn{1}{c}{95.58}          & 95.58          \\ \hline
\multicolumn{1}{c|}{Fine-tuned SAM \& XMem++~\cite{bekuzarov2023xmem}}   & \multicolumn{1}{c}{\textbf{88.17}} & \multicolumn{1}{c}{\textbf{96.16}} & \textbf{96.16} \\ \hline

\end{tabular}} 
\label{Table2}
\end{table}

\section{Conclusion}
This work presents a binary instrument segmentation approach that leverages an efficiently adapted foundation model. Following the segmentation of the initial frames utilizing the foundation model, we employ a target-tracking model to conduct instrument tracking throughout the complete video. This methodology has demonstrated outstanding performance on publicly available datasets and our ESD dataset. Looking ahead, we plan to integrate the pre-trained model into the software infrastructure of the ESD surgical robot system. This integration is expected to enhance the intelligent perception capabilities of robotic systems, enabling them to operate more efficiently in surgical environments.




\bibliographystyle{IEEEtran}
\bibliography{ref}
\end{document}